\documentclass[superscriptaddress,secnumarabic,nobibnotes,aps,
prd,showpacs,nofootinbib, twocolumn]{revtex4}
\usepackage{amsmath}
\usepackage{eurosym}
\usepackage{amssymb}
\usepackage{graphicx}
\usepackage{color}

\def\be{\begin{equation}}
\def\ee{\end{equation}}
\def\bea{\begin{eqnarray}}
\def\eea{\end{eqnarray}}

\newcommand{\f}[2]{\frac{#1}{#2}}

\begin{document}

\title{Reply to ''Comment on 'Can accretion disk properties observationally
distinguish black holes from naked singularities?'''}
\author{Zolt\'{a}n Kov\'{a}cs}
\email{kovacsz2013@yahoo.com}
\affiliation{Max-Fiedler-Str. 7, 45128 Essen, Germany }
\author{Tiberiu Harko}
\email{t.harko@ucl.ac.uk}
\affiliation{Department of Physics, Babes-Bolyai University, Kogalniceanu Street,
Cluj-Napoca 400084, Romania}
\affiliation{School of Physics, Sun Yat-Sen University, Guangzhou 510275, People's
Republic of China}
\affiliation{Department of Mathematics, University College London, Gower Street, London
WC1E 6BT, United Kingdom}
\author{Shahab Shahidi}
\email{s.shahidi@du.ac.ir}
\affiliation{School of Physics, Damghan University, Damghan 41167-36716, Iran}

\begin{abstract}
In the Comment on "Can accretion disk properties observationally distinguish
black holes from  naked singularities?", by Bertrand Chauvineau, Phys. Rev. D {\bf 98}, 088501
(2018), the author
did show that the metric used in Z. Kov\'{a}cs and T. Harko, Phys. Rev.
D {\bf 82}, 124047 (2010), and initially introduced in K. D. Krori and D.
R. Bhattacharjee, J. Math. Phys. \textbf{23}, 637 (1982) and K. K. Nandi, P.
M. Alsing, J. C. Evans, and T. B. Nayak, Phys. Rev. D \textbf{63}, 084027
(2001), does not satisfy the Einstein gravitational field equations with a
minimally coupled scalar field. In our reply we would like to point out that
this result is actually not new, but it was already published in the
literature. Moreover, a rotating solution that generalizes the Kerr metric
for a nonminimally coupled scalar field does exist. We briefly discuss the nature of
the singularities for the generalized metric, and point out that it can be
used as a testing ground to differentiate black holes from naked
singularities. We also mention the existence of some other typing or
technical errors existing in the literature.
\end{abstract}

\maketitle

In Comment on "Can accretion disk properties observationally distinguish
black holes from naked singularities?" by Bertrand Chavineau \cite{0}, the
author did show that the metric introduced in \cite{3} and \cite{4}, and
used in the paper \cite{2} to perform a comparative study of the accretion
disk properties of rotating naked singularities and Kerr type black holes,
does not satisfy the Einstein field equations with a nonminimally coupled
scalar field as a matter source. The
findings of the Comment are {\it undoubtedly correct, and we
fully agree with them}. However, we would like to first point
out that this result is not new, and it has been already known
for some time, being published first in Ref. \cite{1}. When
discussing the metric of Krori and Bhattacharjee \cite{3}, the
authors of \cite{1} explicitly mention that ''{\it However although
this type of metric has been used in a number of later
articles....one can check that the original metric derived by
Krori and Bhattacharjee does not satisfy the field equations...}''
\cite{1}. Unfortunately, when writing our paper \cite{2} {\it we
were not aware that the results by Krori and Bhattacharjee
\cite{3} and Nandi et al. \cite{4} are erroneous}, and thus we have
adopted their proposed rotating geometries as examples of
metrics that could help in distinguishing observationally
between black hole and naked singularity properties. Of
course we also take full responsibility for not checking carefully these previously published results in the literature.
We would also like to emphasize that {\it the use of a metric
that is not an exact solution of the Einstein field equations
could have serious implications on the validity of the
results of} \cite{2}, from both theoretical and observational point
of view.

On the other hand, a rotating solution of the gravitational
field equations in the framework of the Brans-
Dicke theory,
\begin{align}\label{fe}
R_{\mu \nu }=\frac{\omega }{\phi ^{2}}\nabla _{\mu }\phi \nabla _{\nu }\phi +%
\frac{1}{\phi }\nabla _{\mu }\nabla _{\nu }\phi ,
\end{align}%
and
\begin{align}\label{KGe}
\Box \phi =0,
\end{align}
respectively, with the action given by \cite{BD}
\begin{equation}
S=\int d^{4}x\sqrt{-g}\bigg(\phi R-\frac{\omega }{\phi }\nabla _{\mu }\phi
\nabla ^{\mu }\phi \bigg),
\end{equation}%
where $\phi $ is a scalar field that makes the Newton's gravitational
constant dynamical, was also presented in \cite{1} (a similar solution was obtained earlier in \cite{Krr}). The solution is of the
form
\begin{widetext}
\begin{align}\label{kerr}
ds^2&=(\bar{\Delta}\sin^2\theta)^{-2/(2\omega+3)}\Bigg[-fdt^2-\f{4mar}{\rho}\sin^2\theta dtd\phi+\left(r^2+a^2+\f{2ma^2r}{\rho}\sin^2\theta\right)\sin^2\theta d\phi^2\Bigg]\nonumber\\&+(\bar{\Delta}\sin^2\theta)^{2/(2\omega+3)}\rho\left(\f{dr^2}{\Delta}+d\theta^2\right),
\end{align}
\end{widetext}
where we have defined
\begin{eqnarray}
f(r,\theta ) &=&1-\frac{2mr}{\rho },\quad \rho (r,\theta )=r^{2}+a^{2}\cos
^{2}\theta ,  \notag \\
\Delta (r) &=&r^{2}+a^{2}-2mr,\quad \bar{\Delta}=\frac{\Delta}{m^2}.
\end{eqnarray}%
Note that we should assume $\omega \neq -3/2$. In the above metric $m$ and $a
$ are two arbitrary constants, related to the mass and the angular momentum
of the black hole, respectively. The scalar field can be obtained as
\begin{equation}
\phi =(\bar{\Delta} ^{2}\sin ^{4}\theta )^{1/(2\omega +3)},
\end{equation}%
and it satisfies Eq.~(\ref{KGe}). By using a conformal transformation $%
g_{\mu \nu }\rightarrow \tilde{g}_{\mu \nu }=\Omega ^{2}g_{\mu \nu }$, where $\Omega=1/\sqrt{\phi}$ and a
redefinition of the scalar field given by $\tilde{\phi}=\sqrt{2\omega
+3}\ln\phi$, the metric (\ref{kerr})
becomes \cite{1}
\begin{widetext}
\bea\label{kerr1}
ds^{2}=-fdt^2-\f{4mar}{\rho}\sin^2\theta dtd\phi+\left(r^2+a^2+\f{2ma^2r}{\rho}\sin^2\theta\right)\sin^2\theta d\phi^2
+(\bar{\Delta}\sin^2\theta)^{4/(2\omega+3)}\rho\left(\f{dr^2}{\Delta}+d\theta^2\right).
\eea
\end{widetext}
This metric satisfies the field equations \cite{1}
\be
R_{\mu \nu}=\f12 \tilde{\phi}_{\mu}\tilde{\phi }_{\nu}, \quad \Box \tilde{\phi}=0,
\ee
with the scalar field given by
\be
\tilde{\phi}=\f{2}{\sqrt{2\omega +3}}\ln \left(\bar{\Delta} \sin ^2\theta\right).
\ee

The singularities of the  space-time described by the rotating metric (\ref{kerr}) occur at $\Delta=0$,
and $f=0$ and $\rho=0$, respectively, which gives
\begin{align}
r_{\pm}&=m(1\pm\sqrt{1-a_\star^2\cos^2\theta}),  \notag \\
r_{s,n}&=m(1\pm\sqrt{1-a_\star^2}),
\end{align}
where $a_\star=a/m$. Note that $r_\pm$‌ is the surface of infinite redshift and $r_{s,n}$ are the null surfaces and we always have $r_+\geq r_s$.

The Kretchmann scalar $R_{\mu\nu\rho\sigma}R^{\mu\nu\rho\sigma}$ can be
computed as
\begin{align}
R_{\mu\nu\rho\sigma}R^{\mu\nu\rho\sigma}=\frac{512}{\rho^6(2\omega+3)^4}%
(\bar{\Delta}^2\sin^4\theta)^{-\frac{2\omega+5}{2\omega+3}}g(r,\theta),
\end{align}
where $g(r,\theta)$ is a polynomial in $r$ and $\cos\theta$. One can see
that $\rho=0$ is a curvature singularity, which corresponds to
$r=0$ and $\theta=\pi/2$, and it resmbles a ring-like singularity. In the range $\omega<-3/2$ or $\omega>-1/2$, the $r_{s,n}$ are
the Killing horizons \cite{1}.

In the range $-5/2<\omega<-3/2$, $R(r_{s,n})=0$ and we have no curvature singularity in this case. In the opposite case where $\omega>-3/2$ or $\omega<-5/2$, we have $R(r_{s,n})\rightarrow\infty$ and we have two curvature singularities.

From the above relations, we deduce that the curvature singularities $r_{s,n}$ are covered by the horizon for $\omega<-3/2$ and $\omega>-1/2$. In the case $-3/2<\omega\leq-1/2$ there is no horizon and we have three naked singularities $r=0,r=r_{s,n}$.

The  field equations (\ref{fe}) admit a static black hole solution of the form
\begin{align}
ds^2=&-F^{2/\lambda}dt^2\nonumber\\&+\left(1+\f{B}{r}\right)^4F^{2(\lambda-C-1)/\lambda}\bigg[dr^2+r^2d\Omega^2\bigg],
\end{align}
where $F=(1-B/r)/(1+B/r)$, $\phi=\phi_0 F^{C/\lambda}$, with $C$, $b$ and $\lambda$ are constants related to each other as
\begin{align}
\lambda=\sqrt{(C+1)^2-C\left(1-\f{\omega C}{2}\right)}.
\end{align}
It should be noted that the original paper \cite{BD} had a sign typo on the scalar field (the scalar field was written in the form $\phi=\phi_0 F^{-C/\lambda}$), which was corrected by Brans himself in \cite{BD1}. It is interesting that \cite{BD1} has also a typo in the $(00)$ component of the metric tensor.

A metric similar to the Krori and Bhattacharjee metric \cite{3} was considered in \cite{Ag}, but in does not satisfy the Einstein gravitational field equations in the presence of a massless scalar field.

To conclude, rotating Kerr-like solutions of the gravitational field equations for minimally coupled scalar field do exist. These solutions reduce to the standard Kerr metric of standard general relativity in the limit $\omega\rightarrow\infty$, and they can describe both black hole and naked singularity geometries. Therefore, as suggested in \cite{2}, these metrics are the ideal candidates for the investigation of the  Penrose  conjecture, according to which a cosmic censor who forbids the occurrence of naked singularities does exist in nature. They can also offer a possibility of theoretically and observationally differentiating
rotating naked singularities from Kerr-type black holes through the comparative study of their thin disk accretion
 properties.


\begin{thebibliography}{9}
\bibitem{0} B. Chavineau, Comment on "Can accretion disk properties
observationally distinguish black holes from  naked singularities?", Phys. Rev. D {\bf 98}, 088501
(2018).

\bibitem{3} K. D. Krori and D. R. Bhattacharjee, J. Math. Phys. \textbf{23},
637 (1982).

\bibitem{4} K. K. Nandi, P. M. Alsing, J. C. Evans, and T. B. Nayak, Phys.
Rev. D \textbf{63}, 084027 (2001).

\bibitem{2} Z. Kov\'{a}cs and T. Harko, Phys. Rev. D{\bf 82}, 124047 (2010).

\bibitem{1} J. Sultana and B. Bose, Phys. Rev. D {\bf  92}, 104022 (2015).

\bibitem{BD} C. Brans and R. H. Dicke, Phys. Rev. {\bf 124}, 925 (1961).

\bibitem{Krr} H. Kim, Phys. Rev. D {\bf  60}, 024001 (1999).

\bibitem{BD1} C. Brans, Phys. Rev. {\bf 125}, 2194 (1962).

\bibitem{Ag} A. G. Agnese and M. La Camera, Phys. Rev. D {\bf 31}, 1280 (1985).


\end{thebibliography}
\end{document}